\documentclass[11pt]{article}
\usepackage{amssymb}
\usepackage{amsmath}
\usepackage[all]{xy}
\usepackage{graphicx}

\title{A New Objective Definition of Quantum Entanglement as\\Potential Coding of Intensive and Effective Relations.}

\author{{\sc C. de Ronde}$^{1,2,3}$ and {\sc C. Massri}$^{4,5}$}
\date{}

\usepackage[margin=2cm]{geometry}

\begin{document}

\bibliographystyle{plain}
\maketitle

\begin{center}
\begin{small}
1. Philosophy Institute Dr. A. Korn, University of Buenos Aires - CONICET\\
2. Center Leo Apostel for Interdisciplinary Studies\\Foundations of the Exact Sciences - Vrije Universiteit Brussel\\
3. Institute of Engineering - National University Arturo Jauretche\\
4. Institute of Mathematical Investigations Luis A. Santal\'o, UBA - CONICET\\
5. University CAECE
\end{small}
\end{center}

\bigskip

\begin{abstract}
\noindent In \cite{deRondeMassri18d} it was argued against the orthodox definition of quantum entanglement in terms of {\it pure} and {\it separable states}. In this paper we attempt to discuss how the logos categorical approach to quantum mechanics \cite{deRondeMassri18a, deRondeMassri18b} is able to provide an objective formal account of the notion of entanglement ---completely independent of both {\it purity} and {\it separability}--- in terms of the {\it potential coding} of {\it intensive relations} and {\it effective relations}. We will show how our novel redefinition allows us to provide an {\it anschaulich} content to the notion of entanglement, erasing in this way the ``spookiness'' still present within its orthodox understanding in terms of space-time separated collapse particles.
\end{abstract}
\begin{small}

\end{small}

\newtheorem{theo}{Theorem}[section]
\newtheorem{definition}[theo]{Definition}
\newtheorem{lem}[theo]{Lemma}
\newtheorem{met}[theo]{Method}
\newtheorem{prop}[theo]{Proposition}
\newtheorem{coro}[theo]{Corollary}
\newtheorem{exam}[theo]{Example}
\newtheorem{rema}[theo]{Remark}{\hspace*{4mm}}
\newtheorem{example}[theo]{Example}
\newcommand{\proof}{\noindent {\em Proof:\/}{\hspace*{4mm}}}
\newcommand{\qed}{\hfill$\Box$}
\newcommand{\ninv}{\mathord{\sim}} %involutive negation
\newtheorem{postulate}[theo]{Postulate}

\bigskip

\bigskip

\section{Entanglement: Under the Shadow of `Outcomes' and `Particles'}

At the beginning of the 1960s Karl Popper, one of the major figures within the field of epistemology and philosophy of science wrote in {\it Conjectures and Refutations} \cite{Popper62} that the realist account of physics ---the idea that physical theories make reference to {\it physis} (nature or reality)--- had been defeated: ``Today the view of physical science founded by Osiander, Cardinal Bellarmino, and Bishop Berkeley, has won the battle without another shot being fired. Without any further debate over the philosophical issue, without producing any new argument, the {\it instrumentalist} view (as I shall call it) has become an accepted dogma. It may well now be called the `official view' of physical theory since it is accepted by most of our leading theorists of physics (although neither by Einstein nor by Schr\"odinger). And it has become part of the current teaching of physics.'' Popper argued in his book that there existed two main reasons for having reached this difficult situation. The first reason was the impact of Bohr's interpretation of Quantum Mechanics (QM) in the physics community. ``In 1927 Niels Bohr, one of the greatest thinkers in the field of atomic physics, introduced the so-called {\it principle of complementarity} into atomic physics, which amounted to a `renunciation' of the attempt to interpret atomic theory as a description of anything.'' But while pointing his finger to Bohr, the neo-Kantian, Popper seems to have overlooked the deep influence of his own tradition within the instrumentalist path. In fact, logical and empirical positivism ---just like Bohr--- had also shifted the center of analysis from the theory to the subject, from thought to observation, from explanation to prediction. Positivism was born from a riot against Kantian metaphysics and its {\it a priori} concepts. The revolution had begun through an attack commanded by Ernst Mach to the Kantian {\it forms of intuition} founded explicitly in Newtonian classical space and time. The criticisms produced by positivism to Newton's atomist metaphysics at the end of the 19th Century might be regarded as the very condition of possibility for the creation of the two main physical theories of the 20th Century, namely, Relativity and QM. However, Mach's ideas regarding physics ---which understood the discipline as an economy of experience (or observations)---, even though allowed to expand the limits of the theoretical consideration of experience ---beyond classical space-time--- also contained within the seed of instrumentalism itself. The second reason mentioned by Popper for the triumph of instrumentalism makes explicit use of a pragmatic viewpoint stressing the ``the spectacular practical success of [quantum mechanical] applications.'' According to Popper: ``Instead of results due to the principle of complementarity other and more practical results of atomic theory were obtained, some of them with a big bang. No doubt physicists were perfectly right in interpreting these successful applications as corroborating their theories. But strangely enough they took them as confirming the instrumentalist creed.''  However, one might also argue in favor of instrumentalism that what they actually proved ---specially in the influential Manhattan project--- was that spectacular instrumental applications ---such as the atomic bomb--- could be developed without really entering philosophical debates about the meaning and reference of the theory.  

The deep influence of positivism within QM cannot be underestimated. Maybe one the most important points of this decisive influence ---apparently overlooked within the literature--- was the {\it ad hoc} introduction of the {\it projection postulate}. This postulate was required in order to support one of the main cornerstones of positivism, namely, the idea that `physical theories describe the observations of subjects (or agents)'. The famous postulate, interpreted as a ``collapse'' of the quantum wave function, scrambled the objective theoretical representation together with the subjective observation of single `clicks' in detectors. In turn, this addition ---completely foreign to the quantum formalism--- ended up creating the infamous measurement problem in which the literature has been stuck for almost a century \cite{deRonde17a}. But maybe the strangest aspect of this story ---a kind of joke from destiny--- was that, turning against itself, very soon positivism accepted as a self evident ``common sense'' {\it given} the atomist space-time Newtonian metaphysics it had fought and defeated just a few decades before. 

The scrambling of operational definitions with atomist metaphysics within quantum theory is just one exposure ---between many--- of the general confusion which has been systematically reached by mixing objective and subjective forms of analysis and representation. As Jaynes made the point:
\begin{quotation}
\noindent {\small``[O]ur present [quantum mechanical] formalism is not purely epistemological; it is a peculiar mixture describing in part realities of Nature, in part incomplete human information about Nature ---all scrambled up by Heisenberg and Bohr into an omelette that nobody has seen how to unscramble. Yet we think that the unscrambling is a prerequisite for any further advance in basic physical theory. For, if we cannot separate the subjective and objective aspects of the formalism, we cannot know what we are talking about; it is just that simple.''  \cite[p. 381]{Jaynes}}
\end{quotation}

\noindent Just like in the case of other notions used within the theory of quanta, entanglement did not escape this chaotic situation. As it was argued extensively in \cite{deRondeMassri18d}, the metaphysical prejudices regarding the reference of the theory combined with the emphasis on prediction over theoretical representation also reached in the year 1935 Schr\"odinger's definition of {\it quantum entanglement}. Firstly, through the presupposition ---grounded on Newtonian space-time atomism--- that QM makes reference to ``small particles'' and consequently, that an analysis which considers space-time {\it separability} as a standpoint is justified. And secondly, through the ---implicit--- application of the positivist idea according to which QM makes reference to certain (actual) predictions of measurement outcomes (or observations) and consequently, that the notion of ``collapse'' and ``pure state'' should be considered as fundamental cornerstones of any definition of entanglement. This controversial construction of entanglement founded on very weak (metaphysical) pillars added in an {\it ad hoc} manner and unrelated to the mathematical formalism has lead to many difficulties. Indeed, as we discussed in detail in \cite{deRondeMassri18d}, the definition of entanglement is purely negative, it is constructed in contraposition to the notion of {\it separability}. That which is not separable is entangled. In orthodox textbook QM the notion of separability is mathematically represented in terms of the product of subspaces and it is assumed that such product represents naturally the interaction between quantum systems (see e.g., \cite{Hughes92, Jauch68}). However, it is far from clear what is the physical meaning of factorizability in QM, or what is its  operational counterpart. Regardless of its widespread application, factorizability of Hilbert spaces has nothing to do with separability. The product of subspaces simply cannot be interpreted as the addition of `systems' ---as in classical physics. The reason is simple, the product of two rays ($H_1 \otimes H_2$) {\it generate} a whole plane ($H_{12} = H_1 \otimes H_2$) which obviously contains more points than the ones present in the original rays (see also for a more detailed analysis: \cite{Clifton95, Clifton96}). As it is well known in quantum logic, the conjunction in QM is not the same as the union of sets and quantum systems are essentially non-separable \cite{Aerts84b}. Even worse, as shown in \cite{DelaTorre10}, the notion of entanglement is relative to the choice of the factorization of the Hilbert space. Another problem is that the orthodox definition only works with pure states, but not with mixtures. It is a difficult problem to determine if a mixed state is separable or not \cite{LiQuiao18}. Furthermore, as remarked by Einstein and Schr\"odinger in 1935, there seems to exist within the notion of entanglement a strange scrambling of objective correlations with subjective knowledge ---due to the ``collapse'' of the quantum wave functions (section 2). Finally, it is not really clear what quantum entanglement really means since the ``common sense'' explanation in terms of collapsing particles seems to crash with the orthodox formalism of the theory right from the start. 

Taking distance from its orthodox account, in this paper we attempt to provide an objective definition of quantum entanglement which is grounded on the orthodox formalism of QM alone ---without making any metaphysical presupposition whatsoever. In order to do so, we begin in section 2 by addressing the critical definition of entanglement proposed by Schr\"odinger. We then discuss, in section 3, what is meant by objectivity within physical theories. In section 4 we recall some basic aspects of the logos categorical approach to QM from which we attempt to develop our own objective definition of entanglement. After discussing, in section 5, the main elements at play in our proposed redefinition; in section 6, we present a new objective account of quantum entanglement based on the introduction of {\it effective relations} and {\it intensive relations}. In section 7 we analyze in what sense our definition can be considered as being objective. Finally, in section 8, we discuss the intuitive content of our proposed definition of quantum entanglement.

\section{The Entanglement of Systems and Subjective Knowledge}

The notion of {\it entanglement} was introduced by Erwin Schr\"odinger in a series of three papers during the years of 1935 and 1936 \cite{Schr35a, Schr35b, Schr36} which continued the critical reflections introduced by the EPR {\it Gedankenexperiment} \cite{EPR}. In the first paper of the series, {\it Discussion of Probability Relations Between Separated Systems}, he begins right from the start by defining the physical meaning of {\it entanglement} in terms of interacting systems.  
\begin{quotation}
\noindent {\small ``When two systems, of which we know the states by their respective representatives, enter into temporary physical interaction due to known forces between them, and when after a time of mutual influence the systems separate again, then they can no longer be described in the same way as before, viz. by endowing each of them with a representative of its own. I would not call that one but rather the characteristic trait of quantum mechanics, the one that enforces its entire departure from classical lines of thought. By the interaction the two representatives (or $\psi$-functions) have become entangled.'' \cite[p. 555]{Schr35b}}
\end{quotation}

\noindent Immediately after, Schr\"odinger continues to explain how to {\it disentangle} the two systems through measurement: 
\begin{quotation} \noindent {\small ``To disentangle them we must gather further information by experiment, although we knew as much as anybody could possibly know about all that happened. Of either system, taken separately, all previous knowledge may be entirely lost, leaving us but one privilege: to restrict the experiments to one only of the two systems. After reestablishing one representative by observation, the other one can be inferred simultaneously. In what follows the whole of this procedure will be called the {\it disentanglement}. Its sinister importance is due to its being involved in every measuring process and therefore forming the basis of the quantum theory of measurement, threatening us thereby with at least a {\it regressus in infinitum}, since it will be noticed that the procedure itself involves measurement.''\cite[p. 555]{Schr35b}}
\end{quotation}

\noindent Schr\"odinger was making reference here to the famous ``spooky action at a distance''  according to which the measurement ---and subsequent collapse--- of one physical quantity in one of the systems seems to influence the sudden appearance of the definite value of the same physical quantity in the other distant system. Making explicit reference to the EPR paper, Schr\"odinger remarks that: 
\begin{quotation}
\noindent {\small ``Attention has recently \cite{EPR} been called to the obvious but very disconcerting fact that even though we restrict the disentangling measurements to one system, the representative obtained for the other system is by no means independent of the particular choice of observations which we select for that purpose and which by the way are entirely arbitrary. It is rather discomforting that the theory should allow a system to be steered or piloted into one or the other type of state at the experimenter's mercy in spite of his having no access to it.''\cite[pp. 555-556]{Schr35b}}
\end{quotation}

Following the reality criteria proposed in the EPR paper, Schr\"odinger also assumed ---implicitly--- that {\it maximal knowledge} had to be understood as {\it certain knowledge}; i.e., as knowledge involving probability equal to unity. As he critically remarks, the astonishing aspect of QM is that when two systems get entangled through a known interaction, the knowledge we have of the parts might anyhow decrease. 
\begin{quotation}
\noindent {\small ``If two separated bodies, each by itself known maximally, enter a situation in which they influence each other, and separate again, then there occurs regularly that which I have just called {\it entanglement} of our knowledge of the two bodies. The combined expectation-catalog consists initially of a logical sum of the individual catalogs; during the process it develops causally in accord with known law (there is no question of measurement here). The knowledge remains maximal, but at the end, if the two bodies have again separated, it is not again split into a logical sum of knowledges about the individual bodies. What still remains {\it of that} may have become less than maximal, even very strongly so.---One notes the great difference over against the classical model theory, where of course from known initial states and with known interaction the individual states would be exactly known."\cite[p. 161]{Schr35a}}
\end{quotation}

\noindent At this point it becomes evident how the projection postulate interpreted as a ``real collapse'' of the quantum wave function ends up scrambling ---just like in the case of the measurement process--- the objective theoretical representation provided by the mathematical formalism and the subjective observation of a particular `click' in the lab. The entanglement of systems and outcomes within the same representation determines then the scrambling of the objective knowledge, related to the theory, and the subjective knowledge, related to the purely epistemic process of measurement.

\section{In Search of Theoretical Objectivity}

Albert Einstein understood very clearly the importance of what he called the ``detachedness'' of particular subjects (or agents) from the objective representation provided by a physical theory. Objectivity in this case did not refer to the observations of measurement outcomes by subjects (or agents), but ---on the very contrary--- to the representation provided by the theory itself.\footnote{It should be remarked that even though this notion of {\it objectivity} relates directly to its original reference to an object as a {\it moment of unity}; it differs drastically from its contemporary mainstream Bohrian and positivist understanding of objectivity as the {\it intersubjective} account of observations by a community of agents or ---even--- as making reference to the way in which subjects collect information \cite{Dieks09, Dieks18}. Making this shift explicit, Richard Healey remarks in a recent paper \cite{Healey18}: ``Quantum theory is taken to be fundamental to contemporary physics in large part because countless measurements have yielded outcomes that conform to its predictions. Experimenters take great care to ensure that each quantum measurement has an outcome that is not just a subjective impression but an objective, physical event.''} As repeatedly stressed by Einstein, the scrambling of an (objective) theoretical representation and the (subjective) account of observations was completely unacceptable to him. The reason is simple: this scrambling implied the idea that nature was not something that could be represented independently of subjects, but on the contrary, something entangled and dependent on the will and decisions of conscious human beings. This might be the most important point of disagreement with Bohr, who had willingly accepted the idea that measurements played an essential role within the theory and repeated to everyone who was willing to listen to him that the most important lesson to be learnt from QM was an epistemological one, namely, that we subjects are not only spectators but also actors in the great drama of (quantum) existence. As recalled by Pauli, Einstein did not agree with this point which drastically changed the fundaments and reference of physics itself. 
\begin{quotation}
\noindent {\small ``[...] it seems to me quite appropriate to call the
conceptual description of nature in classical physics, which
Einstein so emphatically wishes to retain, `the ideal of the
detached observer'. To put it drastically the observer has according
to this ideal to disappear entirely in a discrete manner as hidden
spectator, never as actor, nature being left alone in a
predetermined course of events, independent of the way in which
phenomena are observed. `Like the moon has a definite position'
Einstein said to me last winter, `whether or not we look at the
moon, the same must also hold for the atomic objects, as there is no
sharp distinction possible between these and macroscopic objects.
Observation cannot {\it create} an element of reality like position,
there must be something contained in the complete description of
physical reality which corresponds to the {\it possibility} of
observing a position, already before the observation has been
actually made'." \cite[p. 60]{Laurikainen88}}
\end{quotation}

\noindent The intromission of subjects within the description of nature was also criticized by Spinoza in the XVII Century, Einstein's favorite philosopher. Spinoza had criticized those who believed that man is an empire within an empire. Subjects are {\it within} nature, and regardless of their self-esteem they are not ``special'' existents. Nature simply does not care what human subjects do. In the same spirit, Einstein ---who believed in the God of Spinoza, i.e. in nature--- is quoted by Everett \cite[p. 7]{OsnaghiFreitasFreire09} to have said that he ``could not believe that  a mouse could bring about drastic changes in the universe simply by looking at it''. He also shared his concerns with Sch\"odinger to whom he wrote \cite[p. 39]{Przibram67}: ``Most [physicists] simply do not see what sort of risky game they are playing with reality ---reality is something independent of what is experimentally established.'' 

In line with Einstein, the kernel of representational realism (as presented in \cite{deRonde16b, deRonde17a}) is that physical theories {\it represent} reality through the creation of interrelated mathematical-conceptual schemes which allow us to provide  a quantitive and qualitative meaning and understanding to the unity present within different fields of experience. According to representational realism, physical representation comes before any other consideration regarding the very possibility of physical analysis. In this respect, the notion of `physical object' ---one of the most important creations of thought--- is also part of a specific representation. This physical representation has been developed both formally and conceptually since Aristotle to Newton. An object is not a ``common sense'' given nor a real existent which is waiting to be observed. The notion of object is a theoretical creation which provides the {\it moment of unity} which through counterfactual reasoning and discourse allows us to make sense of multiple physical phenomena. From this standpoint, the meaning of objectivity is intrinsically related to the possibility of creating representations which, on the one hand, unify multiple experiences coherently, and on the other, are independent ---or `detached'--- from the actions and choices of particular subjects. According to this viewpoint, since there is no moment of unity {\it before} providing the conditions of objective representation, science can only begin its analysis of experience from the standpoint of representation itself. Or in other words, any scientific discourse must always presuppose a conceptual account of what is meant by a `state of affairs'. This is not ---at least for the realist--- something ``self evident'' nor part of the ``common sense'' of the layman but the very precondition for understanding phenomena in a scientific manner. It is the recognition of the need of representation which allows science to be critical about its own foundation. In this respect, we might recall Einstein's remark to Heisenberg that: ``It is only the theory which decides what can be observed.'' This, in fact, was according to Einstein, the really significant  philosophical achievement of Kant:  
\begin{quotation}
\noindent {\small``From Hume Kant had learned that there are concepts (as, for example, that of causal connection), which play a dominating role in our thinking, and which, nevertheless, can not be deduced by means of a logical process from the empirically given (a fact which several empiricists recognize, it is true, but seem always again to forget). What justifies the use of such concepts? Suppose he had replied in this sense: Thinking is necessary in order to understand the empirically given, {\it and concepts and `categories' are necessary as indispensable elements of thinking.}'' \cite[p. 678]{Einstein65} (emphasis in the original)}
\end{quotation} 

Willingly or not, we physicists, are always producing our praxis {\it within} a specific representation. Representation is always first, experience and perception are necessarily second. Paraphrasing Wittgenstein's famous remark regarding language, the physical representation we inhabit presents the limits of the physical world we understand.\footnote{Let us notice, firstly, that ``physical'' should not be understood as a {\it given} ``material reality'', but rather as a procedure for representing reality in theoretical ---both formal and conceptual--- terms. And secondly, that the relation between such physical representation and reality is not something ``self evident''. The naive realist account according to which representation ``discovers'' an already ``fixed'' reality is not the only possibility that can be considered. A one-to-one correspondence relation between theory and reality is a very naive solution to the deep problem of relating theory and {\it physis.}} This marks a point of departure with respect to naive empiricism and positivism, which was also stressed by Einstein: 
\begin{quotation}
\noindent {\small ``I dislike the basic positivistic attitude, which from my point of view is untenable, and which seems to me to come to the same thing as Berkeley's principle, {\it esse est percipi.} `Being' is always something which is mentally constructed by us, that is, something which we freely posit (in the logical sense). The justification of such constructs does not lie in their derivation from what is given by the senses. Such a type of derivation (in the sense of logical deducibility) is nowhere to be had, not even in the domain of pre-scientific thinking. The justification of the constructs, which represent `reality' for us, lies alone in their quality of making intelligible what is sensorily given.'' \cite[p. 669]{Einstein65}}
\end{quotation} 

The naive empiricist viewpoint according to which there can be a ``direct access'' to the world that surround us by ``simply observing what is going on'' was fantastically addressed ---and ironically criticized--- by the Argentine writer Jorge Luis Borges in a beautiful short story called {\it Funes the memorious} \cite{Borges}. Borges recalls his encounter with Ireneo Funes, a young man from Fray Bentos who after having an accident become paralyzed. Since then Funes' perception and memory became infallible. According to Borges, the least important of his recollections was more minutely precise and more lively than our perception of a physical pleasure or a physical torment. However, as Borges also remarked: ``He was, let us not forget, almost incapable of general, platonic ideas. It was not only difficult for him to understand that the generic term dog embraced so many unlike specimens of differing sizes and different forms; he was disturbed by the fact that a dog at three-fourteen (seen in profile) should have the same name as the dog at three fifteen (seen from the front). [...] Without effort, he had learned English, French, Portuguese, Latin. I suspect, however, that he was not very capable of thought. To think is to forget differences, generalize, make abstractions. In the teeming world of Funes there were only details, almost immediate in their presence.''\footnote{The problem exposed by Borges is in fact, the same problem which positivists like Carnap, Nagel, Popper between many others tried to resolve without any success: the difficult relation between, on the one hand, phenomenological experience or observations, and on the other, concepts and theories. An interesting detailed and historical recognition of the many failures of this positivist project is \cite[Chap. 8]{Hempel65}.} Using the story as a {\it Gedankenexperiment} Borges shows why, for a radical empiricist like Funes, there is no reason why to assume a metaphysical identity between `the dog at three-fourteen (seen in profile)' and `the dog at three fifteen (seen from the front)'. For Funes ---the radical empiricist capable of apprehending experience beyond conceptual presuppositions--- there is no `dog', simply because experience does not contain the {\it moment of unity} required to make reference to {\it the same} through time. ``Locke, in the seventeenth century, postulated (and rejected) an impossible language in which each individual thing, each stone, each bird and each branch, would have its own name; Funes once projected an analogous language, but discarded it because it seemed too general to him, too ambiguous. In fact, Funes remembered not only every leaf of every tree of every wood, but also every one of the times he had perceived or imagined it.'' Existence, identity, non-contradiction, are ontological principles which provide the conceptual architectonic which allows us to connect the `the dog at three-fourteen (seen in profile)' and `the dog at three fifteen (seen from the front)' in terms of a {\it sameness}. It is only through these principles that we can think in terms of space-time systems ---such as, for example, a `dog'. Borges shows why  these principles are not self-evident {\it givens} of experience, and neither is a `dog'. And this is the reason why Borges also suspected that Funes ``was not very capable of thought.'' 

Going now back to quantum theory, the question that interests us is the following: {\it Is it possible to provide an objective account of the orthodox quantum formalism?} Objective here should not be understood as ``a true reference to experience'' (whatever that means), but as the formal-conceptual preconditions under which we can represent a {\it moment of unity}, both in the mathematical level and the conceptual discursive level, within a physical theory. It is, in fact, this moment of unity which in turn allows for the famous {\it detachment} of the subject from the represented state of affairs. As it is argued in \cite{deRondeMassri16}, in the case of the mathematical formalisms of physical theories the objective elements are given in relation to the notion of {\it invariance}. If we consider the invariant structure of the quantum formalism it is very easy to detect the ground for any objective consideration: it is obviously the Born rule \cite{deRonde16a, deRondeMassri18a}. What is not so easy of course, is to leave aside the deeply grounded (metaphysical) understanding of physical reality in terms of `systems' and `properties' inhabiting space-time and the actual realm of existence. According to our viewpoint, this actualist representation of physical reality ---when attempting to discuss and analyze what QM is really talking about--- has played the role of an {\it epistemological obstruction}  \cite{deRondeBontems11}. But if we stop trying systematically to relate QM to binary existence, to systems and properties, if we take seriously what the mathematical formalism is telling us, then we must accept ---following the footsteps of Wolfgang Pauli--- that the theory is pointing to the very reconsideration of the way in which we must {\it represent} physical reality in the context of the theory of quanta.

\section{Logos Categorical QM: A Formal-Conceptual Approach}

The logos approach to QM presented in \cite{deRondeMassri18a, deRondeMassri18b} is able to explain in a visualizable manner through the use of graphs, how the objective character of the mathematical representation is restored when replacing the orthodox partial {\it binary valuations} by a {\it global intensive valuation}. By introducing this new type of existential quantification ---grounded on the invariance of the Born rule--- we were able, in turn, to derive a {\it Non-Contextuality Theorem} which shows how to escape Kochen-Specker contextuality and restore an objective reading of the mathematical formalism. But our approach is not only focused in the orthodox mathematical Hilbert scheme, it also stresses the need to supplement the formalism with {\it adequate physical concepts} which must be able to provide an {\it anschaulich} content and explanation of quantum phenomena.\footnote{The {\it anschaulich} aspect of physical theories was something repeatedly discussed by the founding fathers  of QM. More recently David Deutsch, taking distance from empiricists viewpoints which argue that theories are created from observations, has also stressed the importance of their explanatory aspect \cite{Deutsch04, Deutsch16}.} It is by developing new (non-classical) notions that we hope to explain in a new light the basic features already exposed by the formalism of QM. But before addressing the notion of {\it entanglement} more in detail from the perspective of the logos approach to QM, let us first provide a general introduction to its basic elements.

\subsection{Beyond Actuality: Intensive Physical Reality and Objective Probability}  

As it was discussed in  \cite{deRondeMassri18a, deRondeMassri18b}, one of the main standpoints of the logos approach to QM is that by developing a notion of physical reality beyond binary existence ---imposed by the classical representation in terms of `systems', `states' and `properties'---, it is in fact possible to provide a coherent objective representation of QM. This can be done, without changing the orthodox Hilbert formalism, without creating many unobservable worlds or introducing human consciousness within the analysis. According to this viewpoint, there is a main hypothesis presupposed within EPR's line of reasoning and argumentation which is wrong in a fundamental manner. QM simply does not describe an actual {\it separable} state of affairs. And it is the formalism itself which makes explicit this fact ---in many different ways--- right from the start: Heisenberg's indeterminacy principle, the superposition principle, Kochen-Specker theorem, Gleason's theorem, Born's probability rule, they are all ``road signs'' ---as Pauli used to call them--- that point in the direction of leaving behind the classical actualist representation of physics in order to understand the theory of quanta.

It is argued today that physics can only describe `systems' with definite `states' and `properties'. This encapsulation of reality in terms of the classical paradigm ---mainly due to Bohr's philosophy of physics supplemented by 20th Century positivism--- has blocked the possibility to advance in the development of a new conceptual scheme. Taking distance from the Bohrian prohibition to consider physical reality beyond the theories of Newton and Maxwell,\footnote{A prohibition which David Deutsch \cite{Deutsch04} has rightly characterized as ``bad philosophy'', namely, ``[a] philosophy that is not merely false, but actively prevents the growth of other knowledge.''} we have proposed the following extended definition of what can be naturally considered ---by simply taking into account the mathematical invariance of Hilbert formalism--- as a generalized element of (quantum) physical reality (see \cite{deRonde16a}).

\smallskip
\smallskip

\noindent {\it {\bf Generalized Element of Physical Reality:} If we can predict in any way (i.e., both probabilistically or with certainty) the value of a physical quantity, then there exists an element of reality corresponding to that quantity.}

\smallskip
\smallskip

\noindent As it will become clear, this redefinition implies a deep reconfiguration of the meaning of the quantum formalism and the type of predictions it provides. It also allows us to understand Born's probabilistic rule in a new light; not as providing information about a (subjective) measurement result, but instead, as providing objective intensive information of a theoretically described (potential) state of affairs. Objective probability does not mean that particles behave in an intrinsically random manner, it means that probability characterizes an intensive feature of the conceptual representation provided by the quantum formalism accurately and independently of any subjective choice (see also \cite{deRondeFreytesSergioli19}). This account of probability allows us to restore a representation in which the state of affairs is detached from the observer's choices to measure (or not) a particular property ---just like Einstein desired.\footnote{As recalled by Pauli \cite[p. 122]{Pauli94}: ``{\it Einstein}'s opposition to [quantum mechanics] is again reflected in his papers which he published, at first in collaboration with {\it Rosen} and {\it Podolsky}, and later alone, as a critique of the concept of reality in quantum mechanics. We often discussed these questions together, and I invariably profited very greatly even when I could not agree with {\it Einstein}'s view. `Physics is after all the description of reality' he said to me, continuing, with a sarcastic glance in my direction `or should I perhaps say physics is the description of what one merely imagines?' This question clearly shows {\it Einstein}'s concern that the objective character of physics might be lost through a theory of the type of quantum mechanics, in that as a consequence of a wider conception of the objectivity of an explanation of nature the difference between physical reality and dream or hallucination might become blurred.''} This means that within our account of QM ---contrary to the orthodox viewpoint---, the Born rule always provides complete knowledge of the state of affairs described quantum mechanically; in cases where the probability is equal to 1 and also in cases in which probability is different to 1.\footnote{We might talk here of a shift from a {\it binary} understanding of certainty to an {\it intensive certainty}.} As a consequence, all density operators (pure or not) will provide {\it maximal complete knowledge} of a (quantum) state of affairs. Since there is no essential mathematical distinction between states, both pure and mixed states have to be equally considered; none of them being ``less real'', or ``less well defined'' than the other.

\subsection{The Logos Categorical Formalism}

Let us now recall  some basic mathematical notions of our logos categorical approach. We assume that the reader is familiar with the definition of a \emph{category}. Following \cite{deRondeMassri18a}, let $\mathcal{C}$ be a category and let $C$ be an object in $\mathcal{C}$. Let us define the category over $C$ denoted $\mathcal{C}|_C$.
Objects in $\mathcal{C}|_C$ are given by arrows to $C$, $p:X\rightarrow C$,  $q:Y\rightarrow C$, etc. Arrows $f:p\rightarrow q$
are commutative triangles,
\[
\xymatrix{
X\ar[rr]^f\ar[dr]_p& &Y\ar[dl]^q\\
&C
}
\]

\noindent For example, let $\mathcal{S}ets|_\mathbf{2}$ be the category of sets
over $\mathbf{2}$, where $\textbf{2}=\{0,1\}$ and $\mathcal{S}ets$ is
the category of sets.
Objects in $\mathcal{S}ets|_\mathbf{2}$
are functions from a set to $\{0,1\}$
and morphisms are commuting triangles, 
\[
\xymatrix{
X\ar[rr]^f\ar[dr]_p& &Y\ar[dl]^q\\
&\{0,1\}
}
\]
In the previous triangle, $p$ and $q$ are objects of 
$\mathcal{S}ets|_\mathbf{2}$
and $f$ is a function satisfying $qf=p$.

Our main interest is the category $\mathcal{G}ph|_{[0,1]}$ of graphs over the interval $[0,1]$. The category $\mathcal{G}ph|_{[0,1]}$ has very nice categorical properties \cite{quasitopoi, graphtheory}, and it is a \emph{logos}.  Let us begin by reviewing some properties of the category of graphs. A \emph{graph} is a set with a reflexive symmetric relation. The category of graphs extends naturally the category of sets and the category of aggregates (objects with an equivalence relation). A {\it set} is a graph without edges. An {\it aggregate} is a graph  in which the relation is transitive. More generally, we can assign to a category a graph, where the objects are the nodes of the graph and there is an edge between $A$ and $B$ if $\hom(A,B)\neq\emptyset$.
Given that in a category we have a composition law, the resulting graph is an aggregate.

\begin{definition}
We say that a graph $\mathcal{G}$ is \emph{complete} if there is an edge between two arbitrary nodes. A \emph{context} is a complete subgraph (or aggregate) inside $\mathcal{G}$. A \emph{maximal context} is a context not contained properly in another context. If we do 
not indicate the opposite, when we refer to contexts we will be implying maximal contexts.
\end{definition}

\noindent For example, let $P_1,P_2$ be two elements of a graph $\mathcal{G}$. 
Then, $\{P_1, P_2\}$ is a contexts if $P_1$ is related to $P_2$, $P_1\sim P_2$. Saying differently, if there exists an edge between $P_1$ and $P_2$. In general, a collection of elements $\{P_i\}_{i\in I}\subseteq \mathcal{G}$ determine a {\it context} if $P_i\sim P_j$ for all $i,j\in I$. Equivalently, if the subgraph with nodes $\{P_i\}_{i\in I}$ is complete. 

Given a Hilbert space $\mathcal{H}$, 
we can define naturally a graph $\mathcal{G}=\mathcal{G}(\mathcal{H})$
as follows. Following \cite{deRondeMassri18a} the nodes are interpreted as {\it immanent powers} and there exists an edge between  $P$ and $Q$ if and only if $[P,Q]=0$. This relation makes $\mathcal{G}$ a graph (the relation is not transitive). We call this relation {\it quantum commuting relation}.

\begin{theo}
Let $\mathcal{H}$ be a Hilbert space and let $\mathcal{G}$
be the graph of immanent powers with the commuting relation given by QM. 
It then follows that: 
\begin{enumerate}
\item The graph $\mathcal{G}$ contains all the contexts. 
\item Each context is capable of generating the whole graph $\mathcal{G}$.
\end{enumerate}
\end{theo}
\begin{proof}
See \cite{deRondeMassri18b}.
\qed
\end{proof}

\smallskip
\smallskip

As we mentioned earlier, an object in $\mathcal{G}ph|_{[0,1]}$ consists in  a map $\Psi:\mathcal{G}\rightarrow [0,1]$, where $\mathcal{G}$ is a graph. Then, in order to provide a map to the graph of immanent powers, we use the Born rule. To each \emph{power} $P\in\mathcal{G}$, we assign through the Born rule  the number $p=\Psi(P)$, where $p$ is a number between $0$ and $1$ called \emph{potentia}. As discussed in detail in [{\it Op. cit.}], we call this  map $\Psi:\mathcal{G}\rightarrow [0,1]$ a {\it Potential State of Affairs} (PSA for short). Summarizing, we have the following:
\begin{definition}
Let $\mathcal{H}$ be Hilbert space and let $\rho$ be a density matrix.
Take $\mathcal{G}$ as the graph of immanent powers with the quantum commuting relation. 
To each immanent power $P\in\mathcal{G}$ apply the Born rule to get the number $\Psi(P)\in[0,1]$, which is called the potentia (or intensity) of the power $P$.  Then, $\Psi:\mathcal{G}\rightarrow [0,1]$
defines an object in $\mathcal{G}ph|_{[0,1]}$. We call this map a \emph{Potential State of Affairs}.
\end{definition}

Intuitively, we can picture a PSA as a table,
\[
\Psi:\mathcal{G}(\mathcal{H})\rightarrow[0,1],\quad
\Psi:
\left\{
\begin{array}{rcl}
P_1 &\rightarrow &p_1\\
P_2 &\rightarrow &p_2\\
P_3 &\rightarrow &p_3\\
  &\vdots&
\end{array}
\right.
\]

\noindent The introduction of {\it intensive valuations} allows us to derive a non-contextuality theorem that is able to escape Kochen-Specker contextuality [{\it Op. cit.}]. In turn, our approach can be also generalized in a natural manner to density operators. 
\begin{theo}
The knowledge of a PSA $\Psi$ is equivalent to the knowledge of the density matrix $\rho_{\Psi}$. In particular, if $\Psi$ is defined by a normalized vector $v_{\Psi}$, $\|v_{\Psi}\|=1$, then we can recover the vector from $\Psi$.
\end{theo}
\begin{proof}
See \cite{deRondeMassri18b}.
\qed
\end{proof}
\smallskip
\smallskip

Notice that our mathematical representation is objective in the sense that it relates, in a coherent manner and without internal contradictions, the multiple contexts (or aggregates) and the whole PSA. Contrary to the contextual (relativist) Bohrian ``complementarity solution'', there is in this case no need of a (subjective) choice of a particular context in order to define the ``physically real'' state of affairs. The state of affairs is described completely by the whole graph (or $\Psi$), and the contexts bear an invariant (objective) existence independently of any (subjective) choice. Let us remark that `objective' is not understood as a synonym of `real', but rather as providing the conditions of a theoretical representation in which all subjects are {\it detached} from the course of events. Contrary to Bohr's claim, in our account of QM, individual subjects are not considered as actors. Subjects are humble spectators and their choices ---regarding the choice of experimental contexts or mathematical bases--- do not change the objective representation provided by the theory.

\subsection{New Non-Classical Concepts}

Our approach seeks to understand quantum phenomena by introducing notions which are not only capable to match the mathematical formalism in a natural manner, but also allow us to think about what is really going on according to the theory of quanta in a new light. As it is argued in \cite{deRonde16a, deRonde17c, deRonde17a, deRonde17b}, it is the concept of {\it immanent intensive power} which seems particularly well suited to describe what is going on according to the theory of quanta. Let us recall some important features of our new conceptual framework. 

An immanent power, contrary to systems constituted by binary properties, has an {\it intensive existence}. A power is always quantified in terms of a {\it potentia} which measures its strength. The way to compute the potentia of each power is via the Born rule. Due to their invariant character, both powers and potentia can be regarded as being objective, meaning, independent of the (subjective) choice of any particular context. These notions allow us to escape the widespread Bohrian dogma according to which we must simply accept that measurements in QM have a special status.\footnote{It is commonly argued that when we measure in QM, we always influence the quantum object under study ---which is just another way of making reference to the famous ``collapse'' of the quantum wave function. This idea, mainly due to Bohr's account of QM, implies that subjects define reality in an explicit manner; or as Bohr himself used to say: ``that in the great drama of [quantum] existence we are not only spectators but also actors.''} The Born rule is not anymore understood in epistemic terms, as making reference to the probability of obtaining measurement outcomes. Instead, Born's rule is now conceived as a way of computing an objective feature of the (potential) state of affairs represented by QM. The rule provides a definite value of the potentia of each power. The {\it immanent cause} allows us to argue that the single outcome found in a measurement does not influence in any way the superposition as a whole. There is no collapse, no physical process taking place when an agent observes the result of a measurement process. Instead, the finding of an outcome is simply the path from an ontological description to an epistemic inquiry common to all physical theories (see \cite{deRonde17c}). The intensive account of powers also allows us to escape Kochen-Specker type contextuality which becomes in our scheme a purely epistemic feature, one that deals with the {\it epistemic incompatibility} of quantum experiments, and not with the {\it ontic incompatibility} of quantum existents (see \cite{deRonde16c, deRondeMassri18a}). 

As in any other physical theory, within the logos approach, the consideration of quantum measurements must be discussed independently of the mathematical formalism. The incomprehension of QM begun when the notion of measurement was explicitly introduced within the mathematical representation of the theory as an axiom (i.e., the projection postulate). It is this scrambling which has lead QM into all sorts of paradoxes and inconsistencies. Above all, it is responsible for having created the infamous measurement problem; a problem which has influenced in a decisive manner all the research related to the theory of quanta.

\section{A New Objective Path for Quantum Entanglement}

Taking into account right from the start the well known non-classical features of the orthodox quantum formalism, we might ask the following rhetorical questions: 
\begin{itemize}
\item  Why should we expect QM to be related to the classical representation inherited from space-time Newtonian mechanics and the metaphysics of (unobservable) atoms? 
\item Why should we keep using classical notions such as `system', `state' and `property', which we already know very well don't work at all in order to explain the orthodox mathematical formalism? 
\item Why should we accept the naive positivist understanding of observability ---and consequently, the collapse of the quantum wave function--- as a fundamental constraint in order to define physical reality? 
\end{itemize}
In fact, there are many hints coming from the founding fathers which show a different path in order to develop QM beyond the classical (metaphysical) representation of physics. It is by taking seriously both the critical analysis put forward by Einstein and Schr\"odinger and the constructive conceptual approach suggested by Heisenberg's and Pauli's writings, that we have chosen to confront the Bohrian prohibition of developing new conceptual forms imposed by his doctrine of classical concepts and his reductionistic metaphysical presupposition according to which QM must be understood as a rational generalization of classical physics \cite{BokulichBokulich}. Up to now, {\it quantum entanglement} ---following Bohr's restrictions on (classical) language and experience--- has been systematically comprehended in terms of ``interacting elementary particles''. According to the logos approach, what we need to do ---above all--- is to think in a truly different manner. And this can be only done with the aid of a new conceptual framework which ---going beyond classical notions--- allows us to understand what is being observed according to the theory.

\subsection{Beyond Spookiness, Particle Metaphysics and Collapses}

In the history of science, many times, we physicists, have been confronted with ``spooky situations''. The ``spookiness'' always comes from the lack of understanding. Incomprehension of the unknown is always frightening. Just to give an example between many, the phenomena of electricity and magnetism were regarded as ``magical'' since the origin of humanity itself. Pieces of stone attracting each other without any material contact can be indeed ``spooky'', not to talk about lightnings coming from the sky. Until one day physicists were finally able to create a theory called electromagnetism which explained all these different phenomena. Physicists were even able to find out that these apparently different phenomena could be quantified in terms of a unified mathematical formalism. But it was only through the conceptual representation of `fields', that we could finally grasp a deep qualitative understanding of the phenomena. In the end, electricity and magnetism were two sides of the same represented physical reality. Suddenly, the spookiness had disappeared. 

Nobel laureate Steven Weinberg \cite{Weinberg}, when discussing about quantum entanglement, has argued that: ``What is surprising is that when you make a measurement of one particle you affect the state of the other particle, you change its state!'' This conclusion is indeed spooky, since we never observe in our macroscopic world that objects behave in such a strange manner. A table (or chair) in one place never affects the state of another distant table (or chair). If we do something to an object in a region of space $A$, there will be no action produced on an object situated in a distant region $B$. But we must also stress that Weinberg's amazement is implicitly grounded on the assumption that QM talks about ``small particles''. Of course, the atomist Newtonian metaphysical representation of the world has become not only the ``common sense'' of our time but also a very heavy burden for quantum physicists. Advocated like a dogma, without any formal nor experimental support, the idea that QM talks about small particles has remained almost completely unquestioned up to the present. And even though it is not difficult to find researchers who recognize the difficulties of thinking in such classical terms they seem always to forget when talking about the theory in representational terms. But the worst part of this situation comes from those researchers who do not even acknowledge that the Newtonian atomist picture is a particular metaphysical system, and not an ``obvious'' or ``self evident'' way to talk about reality. In fact, it is only when we recognize that physical theories presuppose a formal-conceptual representation that an obvious question raises: could it be possible to create new concepts that would allow us to understand the phenomena implied by {\it quantum entanglement} in a manner which does not consider ``particles'' and which is not ``spooky''? Just in the same way as we created the notion of `field' in order to account for the phenomena of electricity and magnetism, could it be possible to develop a notion which is able to explain in an intuitive manner the correlations that we encounter when performing quantum experiments in the lab?

As we have discussed above, there are two main claims which seem responsible for the ``spookiness'' that we find in the phenomena of entanglement: first, the idea that QM talks about ``small particles''; and second, the idea that superposed particles collapse into single measurement outcomes. While the notion of particle rather than helping understand quantum phenomena seems to have played the role of an epistemological obstruction \cite{deRondeBontems11}, the existence of collapses conflicts with the evolution governed by the Schr\"{o}dinger equation. As remarked by Dieks \cite[p. 120]{Dieks10} regarding the existence of such collapses, ``sophisticated experiments have clearly demonstrated that in interaction processes on the sub-microscopic, microscopic and mesoscopic scales collapses are never encountered.'' In the last decades, the experimental research seems to confirm there is nothing like a ``real collapse'' suddenly happening when measurement takes place. Unfortunately, as Dieks \cite{Dieks18} also acknowledges: ``The evidence against collapses has not yet affected the textbook tradition, which has not questioned the status of collapses as a mechanism of evolution alongside unitary Schr\"odinger dynamics.'' 

At safe distance from many approaches which assume a classical metaphysical standpoint when analyzing QM ---introducing implicitly or explicitly classical notions within the theory---, the logos approach has been devised as a development of the theory which stays close to the quantum formalism in the most strict manner. This implies for us a suspicious attitude towards the (metaphysical) classical notions of `system', `state' and `property'. Taking their place, we have created new non-classical concepts which attempt to satisfy the features of the quantum formalism ---and not the other way around. According to the logos approach, QM does not talk about ``small particles'', it talks about a potential realm ---independent of actuality--- represented in terms of immanent powers with definite potentia. And there is no physical process which takes quantum superpositions into measurement outcomes simply because physical theories represent states of affairs, not observations (or actualities). From this standpoint, we have shown how through the aid of these new concepts it is possible to restore the necessary distance between the objective representation provided by the theory and the subjective measurements taking place {\it hic et nunc} in a lab  \cite{deRonde17c}. The notion of intensive power provides an objective reference to the Born rule \cite{deRonde16a}, escaping the orthodox reading in terms of collapses and measurement outcomes. This new scheme has allowed us to provide an objective and intuitive grasp to the meaning of quantum contextuality \cite{deRondeMassri18a} and quantum superpositions \cite{deRonde17a, deRondeMassri18b}.

\subsection{Beyond Actualist Outcome Coding} 

According to our viewpoint, the main problem with the orthodox understanding of QM is that the representation of physical reality has been completely scrambled with a naive understanding of observability (for a detailed analysis see \cite{deRonde17a}). The infamous ``collapse'' scrambles the objective quantum theoretical representation with subjective epistemic observations. Because of this, the change in our (subjective) knowledge changes the theoretical description of (objective) reality itself.\footnote{This same scrambling takes place in the case of quantum contextuality and the so called ``basis problem''. See for a detailed analysis: \cite{deRonde16c}.} Hence, it is concluded that: ``measurement changes the state of affairs in an uncontrollable manner.'' But, as we have argued extensively, this is not necessarily the end of the story. If we accept that quantum probability is not making reference to measurement outcomes, but instead characterizes an objective feature of the state of affairs represented by QM, then there is no need of considering epistemic measurements within the theoretical representation. In such case, we can restore the objective nature of the phenomena without the intromission of epistemic knowledge within the description of objective quantum reality. The price we have payed willingly is to abandon the atomist metaphysical picture grounded on systems described by definite valued (actual) properties that can be observed with {\it certainty} (i.e., pure states). 

Classically, the coding of a message implies no action at a distance. If we take two envelops and put in one of them a `red' piece of paper, and a `blue' piece of paper in the other one; and we then share the two envelops between two partners which travel to distant places like Buenos Aires and Cagliari; the moment one of the partners in Buenos Aires opens the envelop, he will learn not only which was the paper in his own envelop, he will also learn simultaneously what is the color of the paper within the envelop of his partner in Italy. There is of course nothing spooky about this. None of the pieces of paper traveled from Buenos Aires to Cagliari or vice versa, just because one of the two partners opened the envelop. The state of affairs didn't change. What changed, when opening the envelope, was the knowledge of the state of affairs represented in classical terms. However, this new knowledge did not transform in any way the actual state of affairs represented by classical mechanics. The representation contained the whole set of {\it possible} actual state of affairs. As Einstein and Spinoza demanded, the real situation cannot suddenly change, when a subject chooses to observe what is inside one of the envelopes. The codification of actual correlations must be contained in the theoretical description, and even though the gain of actual knowledge might filter possibilities that did not become actual, it can never change the description provided by the theory. As we shall see in the following section, according to the logos approach, the main difference between classical and quantum codification of information (or entanglement) is not that in the quantum case things are ``spooky'', but that the type of coding is different: while classical entanglement codifies {\it actual binary information}, QM is able to codify {\it potential intensive information}.

\section{Quantum Entanglement as Relational Potential Coding} 

As discussed in the EPR paper, two powers can be related in terms of a {\it definite value}. In his papers, Schr\"odinger also remarks that when considering $X = x_2 - x_1$ or $P = p_2 - p_1$,  $X$ and $P$ might posses a definite value as a whole, say $X'$ and $P'$, but their relata, $x_1, x_2, p_1$ and $p_2$ do not. 
\begin{quotation}
\noindent {\small ``[...] the result of measuring $p_1$ serves to predict the result for $p_2$ and vice versa [since P' is known]. But of course every one of the four observations in question, when actually performed, disentangles the systems, furnishing each of them with an independent representative of its own. A second observation, whatever it is and on whichever system it is executed, produces no further change in the representative of the other system.

Yet since I can predict {\it either} $x_1$ {\it or} $p_1$ without interfering with system No. 1 and
since system No. 1, like a scholar in examination, cannot possibly know which of the two questions I am going to ask it first: it so seems that our scholar is prepared to give the right answer to the first question he is asked, anyhow. Therefore he must know both answers; which is an amazing knowledge, quite irrespective of the fact that after having given his first answer our scholar is invariably so disconcerted or tired out, that all the following answers are `wrong'.'' \cite[p. 555]{Schr35b}}
\end{quotation}

\noindent This relational aspect of QM is not shared by (reductionistic) classical theories grounded on set theory and their underlying Boolean logic. What is difficult to accept from a classical metaphysical viewpoint is that a {\it relation} can have a value without,  at the same time, their {\it relata} possessing one. This first level of relationalism applies to actual observations. The result of the outcome $x_1$ is correlated (or anti-correlated) to the actual outcome $x_2$ (and, of course, viceversa). All this is very well known. However, this is not the only type of relation implied by QM. Quantum relationalism ---addressed in detail in \cite{deRondeFMMassri18a}--- allows a potential coding of powers in two different levels. While in the first level of representation we have {\it definite} or {\it effective relations} dealing with actual measurements, in the second ---most important--- level we have an {\it intensive relation} between powers which requires, at the epistemic level, a statistical type of analysis (see also \cite{deRondeFreytesSergioli19}). It is this intensive level, which we consider to be the most characteristic of QM. Unfortunately, these relations have been overlooked by the physics community embracing the positivist-empiricist obsession towards actual measurement outcomes and their ``common sense'' description in terms of small unobservable particles.

Due to its restricted focus on measurement outcomes, it is only effective relations which have been considered and analyzed by the community discussing quantum information processing. Let us begin by providing a definition of such relations:

\medskip

\noindent {\it {\sc Effective Relations:} The relations determined by a difference of possible actual effectuations. Effective relations discuss the possibility of an actualist definite coding. It involves the path from intensive relations to definite correlated (or anti-correlated) outcomes. They are determined by a binary valuation of the whole graph in which only one node is considered as true, while the rest are considered as false.}

\medskip

\noindent Apart from the actualist coding of measurement outcomes, the potential coding making reference to the potentia of correlated powers must be analyzed in a statistical manner. What needs to be studied in detail is the way in which the potentia of such correlated powers is able to interact in the potential realm. Intensive relations are, according to the logos approach, the true access-key to quantum information processing. 

\medskip

\noindent {\it
{\sc Intensive Relations:} The relations determined by the intensity of different powers. Intensive relations imply the possibility of a potential intensive coding. They are determined by the correlation of intensive valuations.}

\medskip

\noindent In the logos approach it is possible to consider, within a single graph, the entanglement of powers in terms of intensive and effective relations. The quantum situation $\Psi_1|_{C_1}$ exposes, on the one hand, the statistical correlations between two powers, and on the other, the fact that every time we measure we will also obtain correlated effective outcomes. This means that while intensive relations are captured by intensive valuations, effective relations are captured by effective valuations.  

\smallskip

Before we define mathematically the concept of \emph{effective valuation}, 
let us recall the definitions of intensive and 
binary valuations.
An \emph{intensive valuation} (or a PSA) is a map $\Psi:\mathcal{G}\to[0,1]$ 
and a \emph{binary valuation} is a local map $\nu:\mathcal{G}\to\{0,1\}$
both compatible with the structure of the graph. 
Recall that in the quantum case global binary valuations do not exist, 
but global intensive valuations do (see \cite{deRondeMassri18a}).

\begin{definition}
Let $\Psi:\mathcal{G}\to[0,1]$ be a PSA
and let $\mathcal{C}\in \mathcal{G}$ be a context.
An \emph{effective valuation} over $\mathcal{C}=\{P_i\}$ is a random variable $\nu$
which takes $P_k = 1$ and the rest $P_j=0$ (for all $j \neq k$).
\end{definition}
Notice that the concept of effective valuation depends on a context.
Now that we have the mathematical definition of effective valuation, 
let us give the mathematical definition of intensive and 
effective relations.

\begin{definition}
Let $\Psi_1$ and $\Psi_2$ be two PSAs.
We say that $\Psi_1$ and $\Psi_2$ are \emph{related intensively}
if there exists an isomorphism $\tau$ making the following diagram commute
\[
\xymatrix{
\mathcal{G}_1\ar[rr]^{\tau}\ar[dr]_{\Psi_1}& &\mathcal{G}_2\ar[dl]^{\Psi_2}\\
&[0,1]
}
\]
\end{definition}
For example, in the next picture we can visualize the intensive relation
between two PSAs,
\begin{center}
\includegraphics[width=7em]{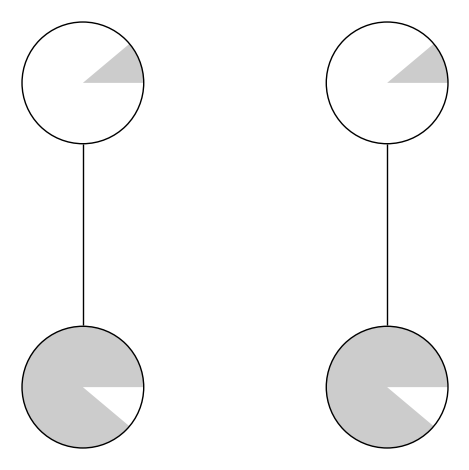}
\end{center}

\begin{example}
Let's give an example that serves for a better understanding of our definitions.
Assume that we have a fair dice with faces $\{1,2,3,4,5,6\}$. 
If we roll this dice on a table, when it comes to rest each face has a probability of $1/6$ of being shown on its upper surface. We can picture this situation in the following way,
\begin{center}
\includegraphics[width=7em]{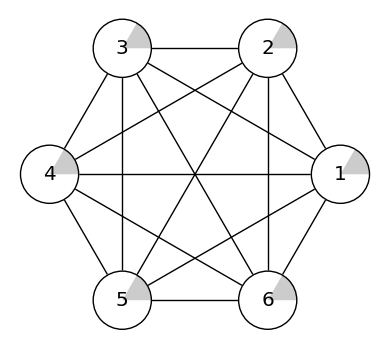}
\end{center}

\noindent Notice that another fair dice, will have the same 
probabilities, hence according to our definitions, both dice will be related intensively. However, the outcome
of the first dice is independent to the outcome of the second one. The intensive relation will not imply an effective relation. 
\qed
\end{example}

\begin{definition}
Let $\Psi_1$ and $\Psi_2$ be two PSAs.
We say that $\Psi_1$ and $\Psi_2$ are \emph{related effectively}
if every effective valuation on $\mathcal{G}_1$ is
correlated (or anti-correlated) to an effective valuation 
on $\mathcal{G}_2$. 
In other words, if any effective valuation $\nu_2$ on $\mathcal{G}_2$
is given by $\tau(\nu_1)$ for some function $\tau$ and some 
effective valuation $\nu_1$ on $\mathcal{G}_1$.
If $\Psi_1$ is 
related effectively to $\Psi_2$, we picture a two-way arrow
between the two graphs.
\end{definition}
For example, the following pictures denote effective relations, 
correlated and anti-correlated,
\begin{center}
\includegraphics[width=7em]{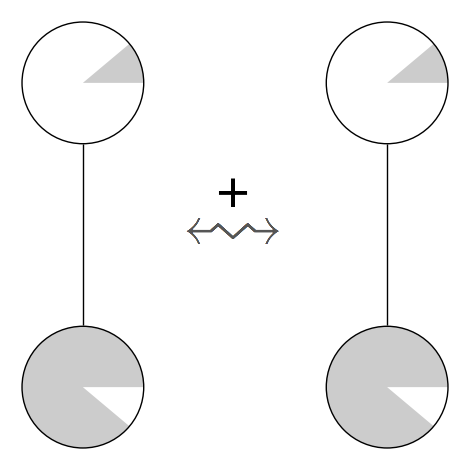}\qquad\qquad
\includegraphics[width=7em]{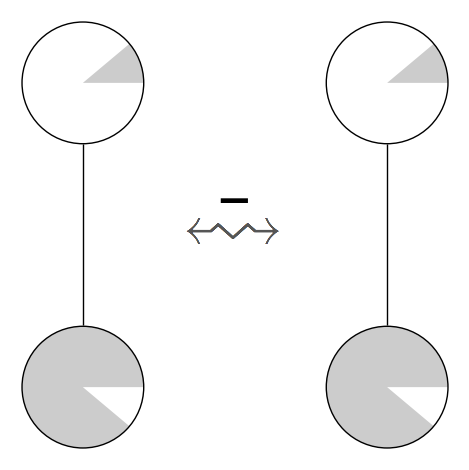}
\end{center}
If the correlation is implicit, we picture effective relations 
in either of the following ways,
\begin{center}
\includegraphics[width=7em]{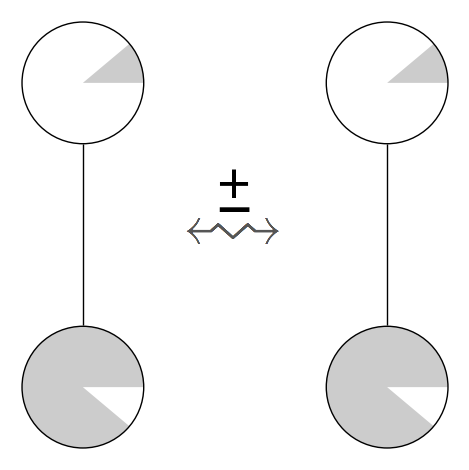}\qquad\qquad
\includegraphics[width=7em]{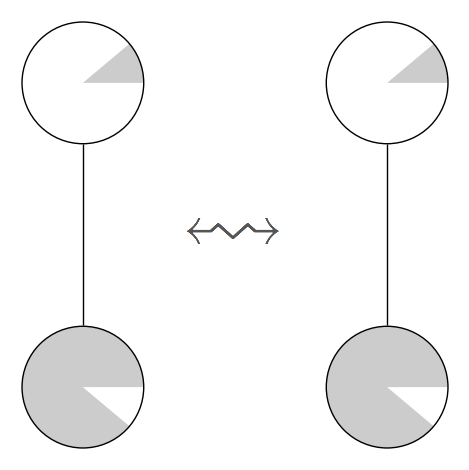}
\end{center}
\begin{example}
Continuing with our example of the two dice, assume that they are both glued together. Then, the outcome of one dice will obviously depend on the outcome of the other one; i.e., they will be related effectively.
\begin{center}
\includegraphics[width=16em]{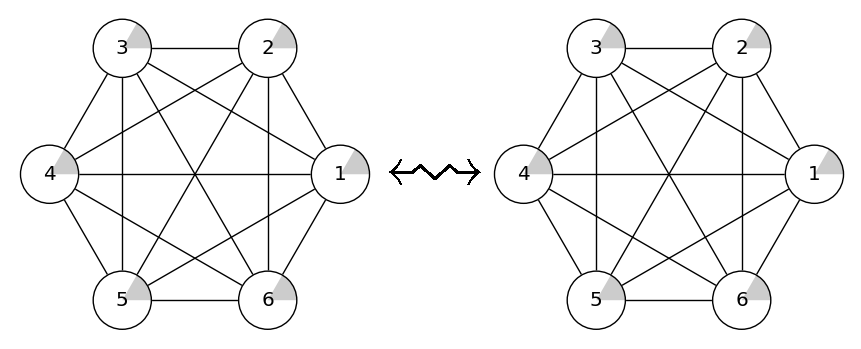}
\end{center}
\qed
\end{example}

\noindent Notice that if there exists an effective relation between two PSAs, then, consequently, there will also exists an intensive relation between them.  But the converse does not follow,
{\small $$\text{EFFECTIVE RELATIONS} \ \ \Rightarrow \ \ \text{INTENSIVE  RELATIONS}$$ 
$$\ \text{INTENSIVE RELATIONS} \ \ \nRightarrow \ \ \text{EFFECTIVE  RELATIONS}$$} 
\noindent Through the notions of intensive and effective relations we are now able to provide a new definition of entanglement.  This definition ---contrary to the orthodox case--- does not depend in any way on the distinction between pure and mixed states nor makes any reference whatsoever to space-time separability.

\begin{definition}[Quantum Entanglement]
Given $\Psi_1$ and $\Psi_2$ two PSAs, if $\Psi_1$ and $\Psi_2$ are related intensively and effectively we say there exists \emph{quantum entanglement} between $\Psi_1$ and $\Psi_2$.
\end{definition}

\noindent This definition of entanglement provides an intuitive grasp related to the correlation between two distant measuring set-ups. We also have the possibility to provide an intuitive non-spacial definition of separability which relates to the lack of correlations between two distant screens.

\begin{definition}[Relational Separability]
Given $\Psi_1$ and $\Psi_2$ two PSAs, if $\Psi_1$ and $\Psi_2$ are not related intensively nor effectively we say there is \emph{relational separability} between $\Psi_1$ and $\Psi_2$.
\end{definition}

It is interesting to notice that our definitions of potential coding in terms of intensive and effective relations allows us to address a third possibility which considers the cases in which there are only intensive relations involved ---but not effective ones. 

\begin{definition}[Intensive Correlation]
Given $\Psi_1$ and $\Psi_2$ two PSAs, if $\Psi_1$ and $\Psi_2$ are related intensively but not effectively we say there exists an \emph{intensive relation} between $\Psi_1$ and $\Psi_2$.
\end{definition}

The ontological account of QM in terms of intensive powers makes explicit that the most important coding of quantum information lies in the intensive statistical level of analysis, not in the level of single measurement outcomes. What needs to be codified is the relational potentia of powers, not effective outcomes. Adding to intensive relations, effective ones show that quantum correlations are not merely ``statistical'' in the classical sense, they are ---as explicitly tested in Aspect's experiments--- much stronger. Unfortunately, restricted by the almost exclusive reference to measurement outcomes, elementary particles and collapses ---a strange mix of instrumentalism and atomist metaphysics---, the orthodox literature has limited itself to the analysis of the actual measurement effectuations of ``two entangled space-time separated particles which collapse each time a measurement is performed''. It is this same picture which has created a ``spooky'' perception about quantum entanglement. As we argued above, it is only a proper conceptual understanding of quantum phenomena which can eradicate this ``spookiness''. For as remarked by Heisenberg:  
\begin{quotation}
\noindent {\small  ```Understanding' probably means nothing more than having whatever ideas and concepts are needed to recognize that a great many different phenomena are part of coherent whole. Our mind becomes less puzzled once we have recognized that a special, apparently confused situation is merely a special case of something wider, that as a result it can be formulated much more simply. The reduction of a colorful variety of phenomena to a general and simple principle, or, as the Greeks would have put it, the reduction of the many to the one, is precisely what we mean by `understanding'. The ability to predict is often the consequence of understanding, of having the right concepts, but is not identical with `understanding'.'' \cite[p. 63]{Heis71}}
\end{quotation}
 What we need to do now, according to the analysis provided by the logos approach, is to advance towards the understanding of the {\it codification of potentia}. This must be done not only in formal categorical terms but also in conceptual representational terms. Only together, the mathematical formalism supported by an adequate conceptual framework will be able to provide a truly objective {\it anschaulich} content to the notion of quantum entanglement.

\section{Restoring an Objective Account of Entanglement}

Any trained physicist knows there is a huge gap between the work done with pen and paper by theoretical physicists and the one done by experimentalists working in the lab with complex instruments. Any student following a theoretical course in physics (e.g., the theory of electromagnetism) has felt the sudden abysm he is confronted with when entering a lab to measure what he just studied in the classroom through complicated equations, the acquisition of new physical concepts and the analysis of ---very far from ``common sense''--- {\it Gedankenexperiments} (e.g., a field). The reason for this huge gap is that the epistemic process of measurement is not contained within the mathematical formalism of any physical theory. Physical theories do not teach us how to actually measure, what instruments are required or how to build a set up. Theories only provide us with a formal-conceptual representation of a state of affairs which can be learnt in a classroom, period. After following a course on classical mechanics one can understand its representation in terms of `interacting particles' in space-time. After following a course on electromagnetism a student will learn how to describe situations in terms of `interacting electro-magnetic waves'. But theoretical courses in physics do not contain any deep technical explanation of how to measure a `particle' or a `field'. Theories simply do not come with a user's manual explaining the technical subtleties of how to measure things in the lab. 

Another important point is that ---contrary to the empiricist claim--- courses always begin with the theory, not with observations nor measurements. As Einstein \cite[p. 175]{Dieks88a} remarked: ``[...] it is the purpose of theoretical physics to achieve understanding of physical reality which exists independently of the observer, and for which the distinction between `direct observable' and `not directly observable' has no ontological significance'' even though, of course, ``the only decisive factor for the question whether or not to accept a particular physical theory is its empirical success.'' It is only the theory which can tell you what can be measured, what the theory talks about. As Heisenberg \cite[p. 264]{Heis73} also made the point: ``The history of physics is not only a sequence of experimental discoveries and observations, followed by their mathematical description; {\it it is also a history of concepts.}  For an understanding of the phenomena the first condition is the introduction of adequate concepts. {\it Only with the help of correct concepts can we really know what has been observed.}'' Physical theories provide an objective representation of a state of affairs detached form the actions and choices of any subject (or agent). Subjects are ---as Einstein correctly stressed--- always detached from theoretical representation. 

It is due to this realist understanding of physics that both Schr\"odinger and Einstein remarked that objective knowledge of a state of affairs cannot change after a known interaction. This is indeed, the very precondition of realism and objective representation. Here, in order to avoid common misunderstandings, the term `knowledge' and the term `interaction' should be clearly specified. The {\it theoretical knowledge} of the objective representation of a state of affairs should not be confused with the {\it epistemic knowledge} acquired by an agent within the lab. Exactly in the same sense, the {\it objective interaction} described by the theoretical representation should not be confused with the {\it actual interaction} that takes place in the lab {\it hic et nunc} when subjects perform measurements. These two levels of analysis ---the objective and the subjective--- should be never scrambled together.  

\medskip

\noindent {\sc Objective Theoretical Knowledge:} The knowledge provided by an objective physical representation. The knowledge provided by the representation is always objective, in the sense that the subject is completely {\it detached} from the represented state of affairs; and complete, in the sense that it cannot be transformed by an epistemic gain of knowledge.

\medskip

\noindent {\sc Subjective Empirical Knowledge:} The knowledge collected by (empirical) subjects through measurements within a lab which requires a specific theoretical representation of the state of affairs under analysis. The epistemic {\it praxis} of measuring a physical quantity is not contained in the mathematical formalism. Epistemic knowledge can be complete or incomplete depending of the technical capabilities and the empirical data collected by a subject (or agent).

\medskip

\noindent {\sc Objective Theoretical Interaction:} An interaction described within a theory. An objective interaction describes the way in which the elements ---described within the theoretical representation--- relate between each other ---{\it within} the representation. For example, a crash of particles as described by classical mechanics or the interaction of two fields according to electromagnetism are cases of objective interactions.

\medskip

\noindent {\sc Subjective Empirical Interaction:} An interaction {\it hic et nunc} of a subject with objects that is not described by a physical theory. For example, the conscious act of perceiving a `click' in a detector, the understanding of what is to be considered a measurement or even the interpretation of what happened.

\medskip

Attempting to unscramble the objective from the subjective in QM, one of the main points of the logos approach is that measurement outcomes (or observations) ---unlike in the orthodox positivist account--- are not considered as part of the theory. Epistemic measurement results are not themselves part of the theoretical representation provided by the orthodox formalism of QM \cite{deRonde16a, deRonde17a, deRondeMassri18b}. By unscrambling the objective theoretical level from the subjective empirical level we've been able to provide an {\it anschaulich} objective account of both contextuality \cite{deRondeMassri18a} and the measurement problem \cite{deRonde17c}. Following this line of research we have provided a new definition of entanglement which restores Einstein's objectivity requirement of a {\it detached observer}. Intensive and effective relations can be addressed without making any reference to measurement outcomes. Given a potential state of affairs, all correlations can be objectively determined via the formalism alone. In this way, quantum entanglement, which is nothing but the correlations which exist within a potential state of affairs, can be defined in a completely objective theoretical manner. Going against the orthodox Bohrian claim according to which our knowledge (or observation) of a quantum system changes the system itself, we have shown that the potential state of affairs does not change when performing measurements. All possibilities of actualization ---as in the case of classical physics--- are already contained within the theoretical description provided by the orthodox formalism of QM.  

%Our claim is that QM provides can provide an objerctive representation of a potential state of affairs in terms of immanent powers with definite potentia. Since we are dealing with an intensive (non-binary) realm of reality, in order to learn about a specific quantum state of affairs present in the lab we always need to perform a repeated statical series of measurements. A single outcome does not provide enough information in order to understand what is going on according to the theory of quanta. In this way, the logos approach is able to introduce a new non-classical account of physical reality which restores the distance between the knowledge provided by an objective theoretical representation and the subjective epistemic knowledge produced by measurements. 

\section{The {\it Anschaulich} Content of Quantum Entanglement} 

From the logos standpoint, we need to answer the following question: What is quantum entanglement in conceptual terms? Or in other words, how can we think intuitively about this physical feature mathematically represented by the quantum formalism in terms of vectors and tensor products? Any adequate account of QM should be able to answer these questions in an intuitive manner. In order to do so, we must abandon the orthodox fictional story of interconnected particles collapsing their states in a super-luminous manner. This story does not make any sense. An adequate conceptual representation should be able to provide an {\it anschaulich} access and the possibility of thinking about the phenomena described by the theory. We believe that the logos approach is able, thorough the addition of new concepts, to produce such a conceptual insight to quantum entanglement. Let us discuss this in some detail. 

We will discuss examples of entanglement provided by two different sports: football ---as we all know it--- and United States' football. Let us begin by football. Argentine Lionel Messi, Brazilian Neymar da Silva Santos J\'unior and Uruguayan Luis Suarez played together in season 2016 at an amazing level. But what was taking place for this to happen? Even though Messi, Neymar and Suarez are three of the best players in the world, there are many examples of very good players which simply don't play well together. This phenomena is well known, the addition of great players doesn't necessarily create a  great team. Indeed, the creation of a football team is a difficult process which is not at all understandable through a reductionistic type of logic. It is not the case that if you have 11 players with the capabilities of Messi, or Diego Armando Maradona, you will necessarily end up with an amazing team. Unlike other sports, the powers of players within a football team are highly contextual and relational. Each position within the field has a very specific need, a particular demand. The powers of a player in one position, say, in the defense, are different from those required in the attack. The aggregate of powers of a player in the left hand side of the field (like Neymar) are different from those playing in the right hand side of the field (like Messi) or even in the middle (like Suarez). Messi is the best player in the world, but in {\it his} specific position ---which, of course, turns out to be one of the most important positions within football due to its proximity to the goal. However, we suspect that Messi would not be such an incredible player acting as a goalkeeper. Messi simply doesn't seem to have the specific powers required in order to act defending the goal in an adequate manner. As we have discussed elsewhere, immanent powers are contextual existents. It is due to this contextual feature of powers that a football team ---as all football fans know--- is something truly difficult to conform. Producing a team implies a process of creating a balanced and capable new individual, namely, the team itself. 

An interesting point for our analysis of quantum entanglement is the way in which the correlations of teams are built up. The process through which a team learns how to play together; or in other words, to ``play as a team''. This means nothing else than to conform actions in a completely correlated manner, as a unity, as an individual. The answer to this question is obvious for anyone who has played a sport: you get more correlated by simply playing together, by training, by interacting through long periods of time. The more a team practices together, the more it will be able to correlate in an adequate manner the relations between the different aggregates of powers the team is composed of. The more a team trains together, the more they interact, the better they will correlate the actual effectuations of powers the day of the game. The interaction between the players creates something which we could also call ``entanglement''; i.e., a potential coding of possible moves of players within the football field. 

The first type of relation is a {\it definite potential coding} which involves the preparation of very specific moves within determined situations. These are called in Argentine ``jugadas preparadas'' ---which translated means ``prepared moves''. Moves prepared in order to deal with corner kicks or free kicks near scoring positions. But there is a much more interesting type of potential correlation to which football fans ---at least in Argentina--- call ``juegan de memoria'', which means that players are able ``to play by heart''. And this is the idea that captures what we mean by {\it intensive potential coding}. Intuitively, it means that some teams are so good in playing together that they act in a completely correlated manner, and this is a kind of non-local behavior. The entanglement of the powers of Messi, Suarez and Neymar gets more correlated the more the players interrelate, the more they practice together, the more they know each other. It is only then, that they can acquire an entangled relationship in which they are able to produce interrelated actions without a previous written plan. For example, when Messi goes to the left, Suarez already knows that he has to go to the right. When Neymar attacks by the left, Messi knows he has to go to the center. The performances that we, football fans, witness every Sunday are highly correlated actions which are not written beforehand. Most of these actions, unlike the case of U.S. football, are not previously determined. The intensive potential coding implies a truly potential correlated aggregate of actions which are not written anywhere. It is like a dance of powers coming into action. And this is why there is a lot of ``instant creation'' within the football field. And also, why football is so interesting to some sensibilities. It is also interesting to notice that the analysis of potential coding does not require effective actualizations. We know that the national team of Argentina and Brazil are better football teams that those of Australia and New Zeland. But this does not mean that Argentina and Brazil will always beat Australia and New Zeland in the field. 

U.S. football and baseball are much more strict regarding the possibilities of creation. The space for the unknown action is much more restricted by the structure of the games themselves. In U.S. football the {\it definite potential coding} of possible moves has a much more important role than in football. In this case the strategy in each situation becomes of outmost importance. This is because U.S. football is not as continuous as football;  it is instead a discrete set of very short specific situations. In each situation, depending on the whole context of the game, the trainer must choose only one between a set of already prepared possible actions. An action is performed and then a new action is required. There is no continuity in the game and the trainer has to become a strategic leader. In this case, the U.S. football player is more a soldier following orders than an artist creating a movement. 

When we go to see a football match, we see how the entanglement encapsulated in potential correlations is actualized during the game. There is nothing spooky about this. As there should be nothing spooky about QM. The potential realm comprises the relations between intensive powers. Potential reality cannot be encapsulated in terms of classical notions. Immanent intensive powers are simply not space-time existents. To end, we might say that against the expectations of Einstein, what is going on cannot be thought within the classical space-time representation; but against Bohrian prohibitions, QM can be explained with non-classical concepts in a truly objective manner.

\section*{Conclusion}

In this paper we provide a new objective definition of the notion of {\it quantum entanglement} from the perspective of the logos categorical approach. Grounded on an intensive interpretation of the Born rule and leaving behind the metaphysics of particles and the existence of collapses, we have provided the definitions of {\it intensive} and {\it effective} relations. Through these notions we are able to characterize a new notion of entanglement in terms of the relational potential coding of intensive and effective relations. We have argued that this definition provides not only an objective account of entanglement but also adds to the {\it anschaulich} understanding of the theory.

\section*{Acknowledgements} 

We want to thank an anonymous referee for her/his insightful remarks and comments. This work was partially supported by the following grants: FWO project G.0405.08 and FWO-research community W0.030.06. CONICET RES. 4541-12 and the Project PIO-CONICET-UNAJ (15520150100008CO) ``Quantum Superpositions in Quantum Information Processing''.

\end{document}